\documentclass[runningheads]{llncs}
\usepackage[T1]{fontenc}
\usepackage{graphicx}
\usepackage{amsmath}

\usepackage{color}
\usepackage{pifont}
\usepackage{algorithm}
\usepackage{algpseudocode}
\usepackage{booktabs}
\usepackage{url}
\usepackage{subfigure}
\usepackage{amsfonts}
\usepackage{booktabs}
\usepackage{bm}
\usepackage{enumitem}
\usepackage{xcolor}
\usepackage{makecell}
\usepackage{bbding}
\usepackage{balance}
\usepackage{microtype}
\usepackage{adjustbox}

\newcommand{\ms}[2]{$#1_{(#2)}$}
\newcommand{\bms}[2]{$\mathbf{#1}_{(#2)}$}
\newcommand{\sbms}[2]{$\underline{#1}_{(#2)}$}

\usepackage{wrapfig}
\usepackage{multirow}
\usepackage{stfloats}
\usepackage{graphicx, subfig}

\usepackage{anyfontsize}

\newcommand{\betweenTinyAndScript}{\fontsize{6.5pt}{8pt}\selectfont}
\usepackage[normalem]{ulem}
\useunder{\uline}{\ul}{}
\usepackage{enumitem}
\begin{document}
\def\method{DSMOE}
\title{Distillation-based Scenario-Adaptive Mixture-of-Experts for the Matching Stage of Multi-scenario Recommendation}


\author{
Ruibing Wang\inst{1} \and
Shuhan Guo\inst{2} \and
Haotong Du\inst{1} \and
Quanming Yao\inst{2}
}

\authorrunning{R. Wang et al.} 

\institute{
Northwestern Polytechnical University, Xi'an, China\\
\email{\{wrb5261, duhaotong\}@mail.nwpu.edu.cn} \and
Tsinghua University, Beijing, China\\
\email{\{guoshuhan, qyaoaa\}@tsinghua.edu.cn}
}
%
%
%
\maketitle              
\begin{abstract}


Multi-scenario recommendation is pivotal for optimizing user experience across diverse contexts. While Multi-gate Mixture-of-Experts (MMOE) thrives in ranking, its transfer to the matching stage is hindered by the blind optimization inherent to independent two-tower architectures and the parameter dominance of head scenarios. To address these structural and distributional bottlenecks, we propose Distillation-based Scenario-Adaptive Mixture-of-Experts (DSMOE). Specially, we devise a Scenario-Adaptive Projection (SAP) module to generate lightweight, context-specific parameters, effectively preventing expert collapse in long-tail scenarios. Concurrently, we introduce a cross-architecture knowledge distillation framework, where an interaction-aware teacher guides the two-tower student to capture complex matching patterns. Extensive experiments on real-world datasets demonstrate DSMOE's superiority, particularly in significantly improving retrieval quality for under-represented, data-sparse scenarios.

\keywords{Recommender Systems  \and Multi-Scenario Matching \and Information Retrieval \and Knowledge Distillation.}
\end{abstract}

\section{Introduction}
With the rapid growth of short-video platforms and e-commerce, \textit{multi-scenario recommendation} has become increasingly important. 
A single application often contains multiple modules or categories, each corresponding to a distinct scenario. 
Users typically maintain stable preferences across contexts (e.g., feeds or content streams), while each scenario introduces unique factors such as layout and interaction style. 
Modeling the interplay between shared preferences and scenario-specific characteristics is thus crucial. 
To this end, MMOE~\cite{ma2018modeling} and its variants (e.g., PLE~\cite{tang2020progressive}, M3OE~\cite{zhang2024m3oe}) have been widely adopted, as their shared experts and scenario-specific gating mechanisms help mitigate negative transfer and balance commonalities with distinctions across scenarios.

Although MMOE-based methods excel in multi-scenario ranking , their application to matching faces the strict latency constraints of billion-scale retrieval. To ensure efficiency, systems typically adopt the Two-Tower architecture~\cite{huang2020embedding}, which encodes user and item representations independently for ANN indexing~\cite{johnson2019billion}. However, this structural independence forces MMOE experts to optimize blindly without early user-item interactions. Furthermore, shared experts are prone to domination by data-rich scenarios, suppressing the modeling of long-tail scenarios , while dedicating experts to each scenario causes prohibitive parameter explosion. 
Consequently, the main challenge lies in resolving the trade-off between the complex interaction modeling of ranking architectures and the retrieval efficiency of matching frameworks.

To address this, we propose \textbf{D}istillation-based \textbf{S}cenario-Adaptive \textbf{M}ixture-\textbf{o}f-\textbf{E}xperts (DSMOE), a novel framework that adapts MMOE to multi-scenario matching tasks by introducing improvements at both the model and optimization levels. On the modeling side, we incorporate \textbf{S}cenario-\textbf{A}daptive \textbf{P}rojection (SAP) \textbf{M}odulation, a lightweight dynamic-parameter module, to prevent experts from being dominated by large scenarios. On the optimization side, we employ a knowledge distillation framework where a powerful teacher model, which takes joint user-item features as input, provides fine-grained guidance for training the MMOE-based two-tower model. This strategy helps the student model overcome the limitations of independent feature encoding while maintaining the high efficiency required for online serving.

The main contributions of this work are summarized as follows:
\begin{itemize}
    \item We propose a lightweight Scenario-Adaptive Projection (SAP) module and integrate it into an MMOE architecture, enabling scenario-specific parameterization that mitigates data imbalance and prevents shared experts from being dominated by data-rich scenarios.
    \item We propose a distillation framework for multi-scenario matching, where a teacher model with joint user-item features guides the two-tower student model to capture complex interactions.
    \item We conduct comprehensive experiments on two real-world public datasets. The results demonstrate that our proposed DSMOE model consistently outperforms strong baselines, especially in data-sparse, long-tail scenarios.
\end{itemize}

\section{Related Work}

\subsection{Matching Problem of Multi-scenario Recommendation}

Recommendation systems~\cite{bian2023feynman,wu2023coldnas,liu2024knowledge} generally consist of two stages: matching and ranking. This work focuses on the multi-scenario matching task~\cite{xie2020internal,jiang2022adaptive,zhao2023m5,zhang2022scenario}, which aims to efficiently retrieve relevant items for users across diverse contexts from a large candidate pool.  

In practice, two-tower architectures are widely adopted for matching~\cite{huang2020embedding}, where user and item embeddings are learned in parallel and compared through Approximate Nearest Neighbor (ANN) search~\cite{johnson2019billion}, enabling high efficiency in industrial systems. Representative models include DSSM~\cite{huang2013learning}, which maps queries and documents into a shared semantic space, and MIND~\cite{li2019multi}, which employs dynamic routing to capture diverse interests.  
Extending this paradigm, recent multi-scenario matching methods explore how global user preferences interact with scenario-specific variations. For instance, ICAN~\cite{xie2020internal} leverages scenario-aware attention, ADIN~\cite{jiang2022adaptive} introduces separate predictors for different scenarios, SASS~\cite{zhang2022scenario} employs a scenario encoder with transfer modules, and M5~\cite{zhao2023m5} combines gating with mixture-of-experts. 

However, a key limitation of this architecture is that user and item features are encoded independently, without early interaction. This restriction prevents the model from capturing complex, high-order feature relationships, which can limit its overall performance in multi-scenario recommendation tasks.
  
Finally, it is important to distinguish matching from ranking~\cite{shen2021sar,chang2023pepnet,tian2023multi,li2023hamur,zhu2024m,gao2024hierrec}. Ranking focuses on detailed comparisons within a small candidate set, whereas matching must efficiently filter from a much larger pool, making ranking models unsuitable for direct deployment in the matching stage.  



\subsection{The Multi-gate Mixture-of-Experts (MMOE) Backbone}
The Multi-gate Mixture-of-Experts (MMOE) framework~\cite{ma2018modeling} captures scenario-specific patterns through shared experts and scenario-dependent gating networks. Formally, for a scenario $t$, the output is defined as:
\begin{equation*}
    y_t(x) = \sum\nolimits_{k=1}^{K} g_t^{(k)}(x) \cdot f^{(k)}(x),
\end{equation*}
where $f^{(k)}(x)$ denotes the expert output and $g_t^{(k)}(x)$ is the gating weight. 

Several studies have applied MMOE to the ranking stage of multi-scenario recommendation. 
HMoE~\cite{li2020improving} further enhances MMOE through a hierarchical MoE design that models scenario-specific and task-shared representations via layered expert routing. 
HiNet~\cite{zhou2023hinet} further extends MMOE with hierarchical extraction at both the scenario and task levels, combined with explicit cross-scenario attention. 
M3OE~\cite{zhang2024m3oe} integrates multi-domain and multi-task learning by coupling a shared representation space with domain- and task-specialized experts.

However, the Two-Tower architecture restricts expert gating to unilateral user features, preventing the early interaction typical in ranking. Furthermore, shared experts suffer from gradient domination by data-rich scenarios, degrading representations in long-tail contexts

\section{Methodology}


\begin{figure*}[t]
	\centering
	\includegraphics[width=0.99\textwidth]{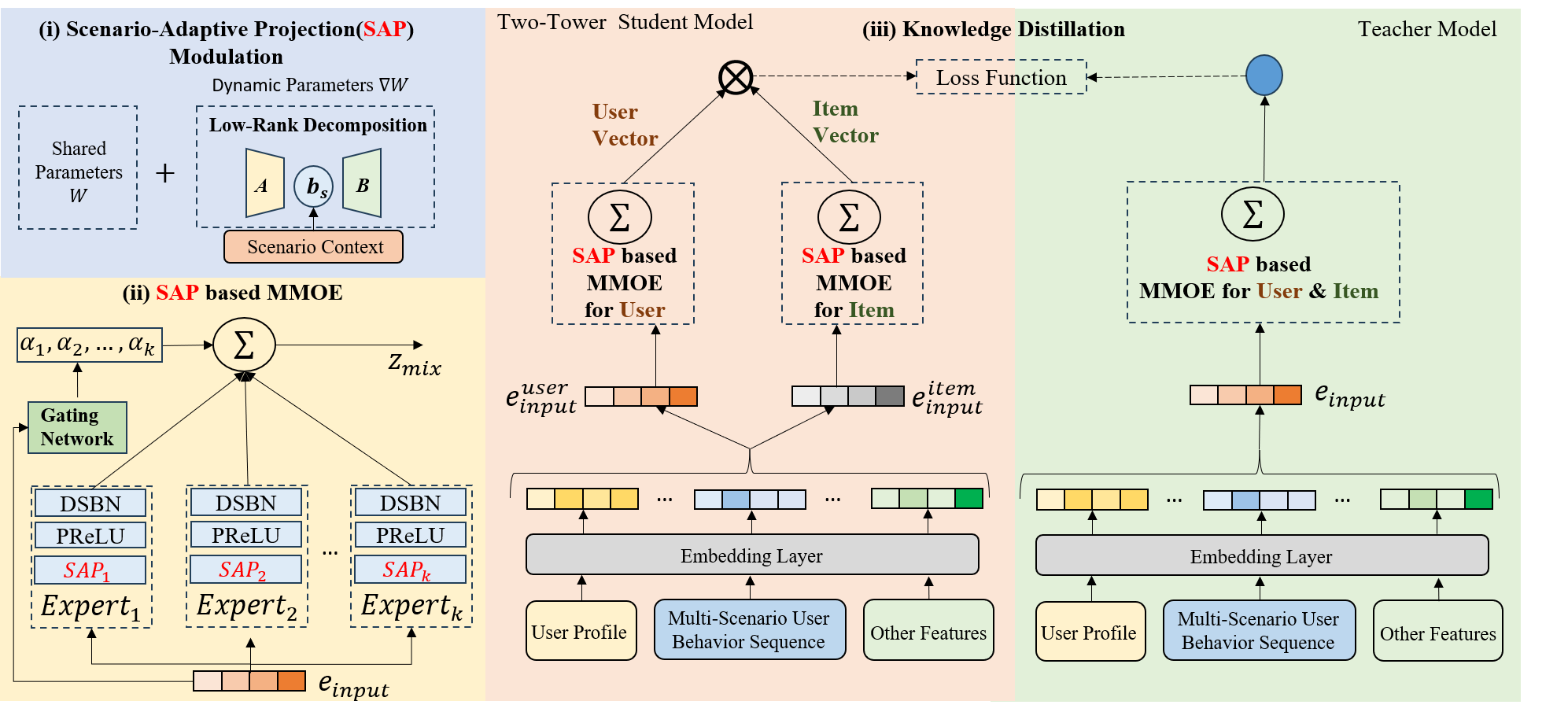}
	\caption{The framework of DSMOE. (i) The \textbf{S}cenario-\textbf{A}daptive \textbf{P}rojection (SAP) module, which uses low-rank decomposition to generate dynamic parameters from scenario context. (ii) The SAP-based MMOE architecture. (iii) The knowledge distillation optimization framework.}
\label{fig:illus}
\end{figure*}

\subsection{Problem Formulation}

We study the \textit{multi-scenario matching} task, whose objective is to retrieve a set of items that align with a user's interests from a large-scale candidate pool, while explicitly accounting for the scenario in which the recommendation occurs.

Let $\mathcal{U}=\{u\}$ denote the set of users, $\mathcal{V}=\{v\}$ the set of items, and $\mathcal{S}=\{s\}$ the set of scenarios. The matching process for a given user--scenario pair $(u,s)$ can be expressed as:
\begin{equation*}
    \mathcal{R}_{u,s} = \underset{v \in \mathcal{V}}{\arg\text{Top}_K} \; g_{\theta}(v \mid u, s),
\end{equation*}
where $\mathcal{R}_{u,s}$ is the set of top-$K$ items returned for user $u$ in scenario $s$, and $g_{\theta}$ is the matching model parameterized by $\theta$.


However, the Two-Tower paradigm imposes a trade-off: its structural independence restricts early interaction, while shared parameters struggle with scenario imbalance. Our goal is to resolve this conflict by imparting high-order interaction capabilities and scenario adaptivity into the efficient Two-Tower architecture.


\subsection{Framework Overview} 

To bridge MMOE's modeling capacity with matching efficiency, we propose the DSMOE framework (Figure 1). We address structural and distributional bottlenecks by incorporating \textit{Scenario-Adaptive Projection Modulation} (Section \ref{sec:Architecture}) to prevent expert domination and employing a teacher model with joint inputs (Section \ref{sec:Optimization}) to guide the Two-Tower student.

\subsection{Model Architecture} \label{sec:Architecture}


\subsubsection{Embedding Layer}

We embed each raw user feature into a $d_{\text{emb}}$-dimensional dense vector. User features are categorized as \textit{sparse} (e.g., user ID, gender), \textit{dense} (e.g., registration days), or \textit{sequential} (e.g., multi-scenario behavior sequences).

For the $m$-th feature $f_{u,m}$ of user $u$, 
its embedding vector $\mathbf{e}_{u,m}$ is obtained as
\begin{align*}\label{eq:emb-layer}
\mathbf{e}_{u,m} =
\begin{cases}
    \mathbf{W}^{(u)}_{m} \cdot \mathrm{onehot}(f_{u,m}), & \text{if $f_{u,m}$ is a sparse feature}, \\
    \mathbf{W}^{(u)}_{m} \cdot f_{u,m}, & \text{if $f_{u,m}$ is a dense feature}, \\
    \mathrm{POOLING}\big(f_{u,m}\big), & \text{if $f_{u,m}$ is a sequential feature},
\end{cases}
\end{align*}
where $\mathbf{W}^{(u)}_{m}$ denotes the learnable embedding or projection matrix for the $m$-th feature field, 
$\mathrm{onehot}(f_{u,m})$ is the one-hot encoding of a sparse feature, 
and $\mathrm{POOLING}(\cdot)$ represents a temporal pooling operation (e.g., max or average pooling) applied over the sequence embeddings.

Scenario features $f_s$ (e.g., scenario indicator) are similarly embedded:
\begin{equation}
\label{eq:e_scenario}
\mathbf{e}_{s} = \mathbf{W}^{(s)} \cdot \mathrm{onehot}(f_s).
\end{equation}

\subsubsection{Scenario-Adaptive Projection Modulation(SAP)}

Dynamic-parameter techniques have been widely applied in the ranking stage of multi-scenario recommendation, but they are not yet commonly used in the matching stage due to efficiency constraints that make them less practical. To address this challenge, we propose \textit{Scenario-Adaptive Projection Modulation} (SAP), a lightweight dynamic-parameter module for multi-scenario modeling (Fig~\ref{fig:illus} (i)). SAP layers are typically built on the basic linear transformation $y = Wx$, by refining this unit with scenario-adaptive optimization:
\begin{equation*}
\label{eq:SAP}
Layer_{SAP} = W_{\text{shared}} x + \big(\Delta W(e_{\text{s}})\big) x ,
\end{equation*}
where $W_{\text{shared}}$ is a base weight matrix shared across all samples, and $\Delta W(e_{\text{s}})$ is a low-rank correction matrix dynamically generated from the scenario feature $e_{\text{s}}$. The generation process is defined as:
\begin{equation*}
\Delta W(e_{\text{s}}) = \sum\nolimits_{r=1}^{R} (b_s)_r \cdot \big(A_{:,r} \, B_{r,:}\big) ,
\end{equation*}

Here, $A \in \mathbb{R}^{d_{\text{in}} \times R}$ and $B \in \mathbb{R}^{R \times d_{\text{out}}}$ are two trainable low-rank decomposition matrices, with $R \ll d_{\text{in}}, d_{\text{out}}$, and $b_s \in \mathbb{R}^R$ is a bias vector generated from the scenario feature as $b_s = \text{Linear}_{\text{s}}(e_{\text{s}})$.

This design assigns each scenario a small set of dynamic parameters, thereby alleviating data imbalance across scenarios.

\subsubsection{SAP based MMOE}

Although MMOE can leverage multiple experts to model multi-scenario tasks, shared experts are often dominated by data-rich scenarios, while scenario-specific experts may suffer from insufficient training in small-scale scenarios. To address this issue, we propose constructing experts with \textit{Scenario-Adaptive Projection (SAP)}, which allocates lightweight scenario-specific parameters to enhance modeling capacity for small-scale scenarios (Fig~\ref{fig:illus} (ii)).


\paragraph{Expert Networks.}
Each expert is implemented as:
\begin{equation*}
\mathrm{Expert}_k(x \mid e_{\text{s}}, d) \;=\; \mathrm{DSBN}^{(k)}\!\big(\mathrm{PReLU}(\mathrm{SAP}^{(k)}(x \mid e_{\text{s}})),\, d\big),
\end{equation*}
where $x \in \mathbb{R}^{d_{\text{in}}}$ is the expert input feature vector, $e_{\text{s}} \in \mathbb{R}^{d_{\text{dom}}}$ is the scenario feature vector obtained by \eqref{eq:e_scenario}, $d$ is the scenario ID, $\mathrm{SAP}^{(k)}(\cdot \mid e_{\text{s}})$ is the scenario-adaptive projection modulation obtained by \eqref{eq:SAP}, $\mathrm{PReLU}(\cdot)$ is the parametric ReLU activation, $\mathrm{DSBN}^{(k)}(\cdot, d)$ applies scenario-specific batch normalization with parameters $(\gamma_d, \beta_d)$ for scenario $d$ in expert $k$.

\paragraph{Scenario-Aware Mixture.}
Given the scenario feature $e_{\text{s}}$, each expert gate computes a logit $g_k = (W^{\text{gate}}_k)^{\top} e_{\text{s}} + b^{\text{gate}}_k$, which is normalized via softmax to obtain weights $\alpha_k = \exp(g_k)/\sum_j \exp(g_j)$. 
The output is then aggregated as 
\begin{equation*}
\mathbf{z}_{\text{mix}} = \sum\nolimits_{k=1}^{K} \alpha_k \cdot \mathrm{Expert}_k\big(\mathbf{e}_{\text{input}} \mid e_{\text{s}}, d\big),
\end{equation*}
where $\mathbf{e}_{\text{input}}$ denotes the concatenated embeddings of the input features.
.

The mixture representation $\mathbf{z}_{\text{mix}}$ is then fed into a forward network, also implemented with the Scenario-Adaptive Projection (SAP) module, to perform the final dimensional transformation and information refinement, yielding the output vector $\mathbf{z}_{\text{output}} = \mathrm{SAP}_{forward}(\mathbf{z}_{\text{mix}})$.





\subsection{Model Optimization} \label{sec:Optimization}

To address the limitation where strict tower separation prevents capturing complex feature interactions , we propose a knowledge distillation framework utilizing a teacher model that processes concatenated features to explicitly model high-order interactions. By optimizing a joint loss, the student model mimics the teacher's outputs, effectively inheriting rich interaction knowledge to mitigate the blind optimization of experts.

We formulate multi-scenario matching as a binary classification task with random negative sampling. 
For a user–item pair $(u,v)$, the matching score is defined as $\hat{y} = \sigma(\langle \hat{\mathbf{e}}_u, \hat{\mathbf{e}}_v \rangle)$, 
where $\langle \cdot, \cdot \rangle$ denotes the inner product and $\sigma(\cdot)$ is the sigmoid function.
The task loss is then defined as
\begin{align*}
\label{eq:loss_task}
\mathcal{L}_{task} = \sum\nolimits_{(u,v,y)\in\mathcal{T}} \Big(y\log \hat{y} + (1-y)\log(1-\hat{y})\Big),
\end{align*}
where $\mathcal{T}$ denotes the training set.

To further enhance supervision, we introduce a knowledge distillation (KD) framework. 
Let $p_{t}$ be the probability predicted by the fused teacher model and $p_{s}=\hat{y}$ be the probability from the student model. 
The overall objective combines the task loss with a distillation term:
\[
\mathcal{L}_{total} = \mathcal{L}_{task} + \lambda \, \mathcal{L}_{KD}(p_{t}, p_{s}),
\]
where $\mathcal{L}_{KD}$ is defined as the Kullback--Leibler divergence $\text{KL}(p_{t} \parallel p_{s})$.

This distillation transfers complex user-item interaction knowledge to the student, providing richer supervisory signals than sparse binary labels. Crucially, the teacher is discarded after offline training; only the lightweight student model is deployed, preserving the high efficiency of the two-tower paradigm.


\vspace{-5pt}
\begin{table*}[h]
\caption{Comparison of DSMOE with existing methods. \textbf{Interaction Awareness} indicates whether the model captures complex user-item interactions beyond simple dot products during training.}
\label{tab:comparison}
\centering
\renewcommand{\arraystretch}{0.85}
\resizebox{0.85\textwidth}{!}{
\footnotesize
\begin{tabular}{lcccc}
\toprule
Method & Stage & Architecture & ANN & Interaction Awareness \\
\midrule
HMOE \cite{li2020improving} & Ranking & Single-Tower & \ding{55} & High (Early Interaction) \\
M3oE \cite{zhang2024m3oe} & Ranking & Single-Tower & \ding{55} & High (Early Interaction) \\
\midrule
ADIN \cite{jiang2022adaptive} & Matching & Two-Tower & \ding{51} & Low (Late Interaction) \\
SASS \cite{zhang2022scenario} & Matching & Two-Tower & \ding{51} & Low (Late Interaction) \\
\midrule
\textbf{DSMOE} & \textbf{Matching} & \textbf{Two-Tower} & \textbf{\ding{51}} & \textbf{High (via Distillation)} \\
\bottomrule
\end{tabular}
}
\end{table*}
\vspace{-5pt}
\subsection{Comparison with Existing Works}
Table \ref{tab:comparison} compares DSMOE with existing methods. Ranking approaches (e.g., HMOE, M3oE) possess high interaction awareness but lack the retrieval efficiency required for ANN indexing. Conversely, matching methods (e.g., ADIN, SASS) ensure efficiency via two-tower architectures yet suffer from low interaction awareness due to simple late fusion. DSMOE bridges this gap by distilling complex interaction signals into an efficient two-tower framework, simultaneously achieving high accuracy and deployment feasibility.

\section{Experiments}
\subsection{Experimental Settings}
\subsubsection{Datasets}\label{sec:dataset}
We conduct experiments on two widely used public datasets.
KuaiRand-Pure~\cite{gao2022kuairand} originates from the interaction logs of Kuaishou, a short-video platform. The scenario context is defined by the attribute tab.
Alimama~\cite{gai2017learning,zhou2018deep} is provided by the Alimama advertising platform. The data is partitioned into four scenarios according to city level. In dataset partitioning, we adopted the settings proposed in PERSCEN~\cite{du2025perscen}.

\subsubsection{Evaluation Metrics}
Following prior studies~\cite{jiang2022adaptive,zhang2022scenario}, 
we adopt Recall@K to assess the quality of candidate matching. 
Recall@K measures the proportion of relevant items retrieved within the top-$K$ results. 
In our experiments, the value of $K$ is chosen to be roughly 1\% of the total candidate set size. 
Specifically, we set $K \in \{50, 100\}$ for KuaiRand-Pure and $K \in \{500, 1000\}$ for Alimama.

\subsubsection{Implementation Details}
All models were implemented in PyTorch and optimized using the Adam optimizer with a learning rate of $1\times10^{-4}$ and a weight decay of $1\times10^{-6}$. We adopted a batch size of 4096 and trained for up to 50 epochs. The embedding dimension for all feature fields was fixed at 16. 
For our proposed \method{} model, the number of experts was set to $K=3$ and the rank of the SAP module was set to $R=4$, as determined by the sensitivity analysis in Section~\ref{app:hyper}. 
All reported results are averaged over five independent runs conducted on an NVIDIA GeForce RTX 3090Ti GPU.

\subsection{Performance Comparison}\label{sec-perf}

We evaluate the proposed \method{} model against 
advanced approaches such as \textbf{ICAN}~\cite{xie2020internal}, \textbf{ADIN}~\cite{jiang2022adaptive}, \textbf{SASS}~\cite{zhang2022scenario}, \textbf{M5}~\cite{zhao2023m5}, and \textbf{PERSCEN}~\cite{du2025perscen}.  


As shown in Table~\ref{tab:results}, DSMOE significantly outperforms all baseline methods on both the KuaiRand-Pure and Alimama datasets. The improvement is especially noticeable in data-sparse scenarios, where standard MMOE-based approaches often struggle because gradients from dominant scenarios overwhelm the others. This demonstrates that our SAP module effectively balances learning across different scenarios and helps preserve useful information for long-tail cases. In addition, the consistent gains in Recall indicate that the distillation process successfully transfers fine-grained user-item interaction signals that are typically used only in ranking models into the shared embedding space, resulting in strong performance across a wide range of data densities.

\begin{table*}[t]
        \betweenTinyAndScript
	\caption{Test performance obtained on KuaiRand-Pure and Alimama. 
		The best results are bolded, the second-best results are underlined. 
		The proportion of each scenario’s data in the overall dataset is shown in $(\cdot)$ after the scenario identifier, highlighting data sparsity. For example, K1 (84\%) indicates that K1 data accounts for 84\% of the KuaiRand-Pure dataset.
	}
        \label{tab:results}
        \resizebox{\textwidth}{!}{%
		\begin{tabular}{c|cc|cc|cc|cc}
			\hline
			\multirow{2}{*}{\textit{KuaiRand-Pure}} & \multicolumn{2}{c|}{K1~(84\%)} & \multicolumn{2}{c|}{K2~(9\%)} & \multicolumn{2}{c|}{K3~(4\%)} & \multicolumn{2}{c}{K4~(3\%)} \\	
			& R@50(\%) & R@100(\%) & R@50(\%) & R@100(\%) & R@50(\%) & R@100(\%) & R@50(\%) & R@100(\%)\\	
			\hline
			ICAN & \ms{18.06}{0.16} & \ms{28.93}{0.15} & \ms{21.63}{0.62} & \ms{31.99}{0.37} & \ms{17.64}{0.29} & \ms{28.19}{0.61} & \ms{3.47}{0.26} & \ms{6.98}{0.32}\\
			ADIN & \ms{18.46}{0.03} & \ms{29.29}{0.25} & \ms{29.50}{0.52} & \ms{40.83}{0.39} & \ms{18.88}{0.62} & \ms{29.82}{1.37} & \ms{19.12}{1.47} & \ms{29.72}{1.71}\\
			SASS & \ms{17.70}{0.18} & \ms{28.28}{0.30} & \ms{30.10}{0.31} & \ms{41.25}{0.43} & \ms{17.99}{0.67} & \ms{28.03}{0.99} & \ms{18.95}{0.78} & \ms{28.17}{1.07}\\
            M5 & \ms{17.91}{0.08} & \ms{28.62}{0.25} & \ms{28.06}{0.47} & \ms{39.55}{0.42} & \ms{18.57}{0.95} & \ms{27.98}{0.65} & \ms{15.81}{1.55} & \ms{24.51}{1.27} \\	
			PERSCEN & \sbms{18.74}{0.13} & \sbms{29.95}{0.12} & \sbms{30.69}{0.21} & \sbms{42.62}{0.30} & \sbms{19.60}{0.86} & \sbms{30.50}{0.89} & \sbms{21.39}{0.83} & \sbms{31.68}{1.11} \\
            \method{} & \bms{18.86}{0.24} & \bms{30.06}{0.23} & \bms{31.04}{0.32} & \bms{42.66}{0.18} & \bms{20.26}{0.75} & \bms{30.61}{0.78} & \bms{21.87}{0.72} & \bms{32.60}{0.64} \\
			\hline \hline
			\multirow{2}{*}{\textit{Alimama}} & \multicolumn{2}{c|}{A1~(45\%)} & \multicolumn{2}{c|}{A2~(25\%)} & \multicolumn{2}{c|}{A3~(20\%)} & \multicolumn{2}{c}{A4~(10\%)} \\	
			& R@500(\%) & R@1000(\%) & R@500(\%) & R@1000(\%) & R@500(\%) & R@1000(\%) & R@500(\%) & R@1000(\%)\\	
			\hline
			ICAN & \ms{9.76}{1.35} & \ms{15.02}{1.08} & \ms{9.81}{1.16} & \ms{15.28}{0.90} & \ms{9.51}{1.24} & \ms{14.91}{0.98} & \ms{10.18}{1.27} & \ms{15.55}{0.90}\\
			ADIN & \ms{11.01}{0.52} & \ms{15.50}{0.62} & \ms{11.38}{0.67} & \ms{16.13}{0.71} & \ms{10.94}{0.34} & \ms{16.02}{0.50} & \ms{11.82}{0.50} & \ms{16.72}{0.38} \\
            SASS & \ms{10.39}{0.14} & \ms{15.06}{0.10} & \ms{10.80}{0.17} & \ms{15.72}{0.20} & \ms{10.36}{0.24} & \ms{15.32}{0.31} & \ms{11.04}{0.58} & \ms{15.86}{0.44} \\
			M5 & \ms{11.01}{0.36} & \ms{15.71}{0.27} & \ms{11.06}{0.17} & \ms{15.56}{0.19} & \ms{10.95}{0.26} & \ms{15.49}{0.44} & \ms{11.00}{0.60} & \ms{15.45}{0.78} \\	
			PERSCEN & \sbms{12.72}{0.20} & \sbms{17.63}{0.37} & \sbms{12.66}{0.32} & \sbms{17.60}{0.22} & \sbms{12.34}{0.24} & \sbms{17.32}{0.14} & \sbms{12.63}{0.12} & \sbms{17.47}{0.42} \\
            \method{} & \bms{12.84}{0.31} & \bms{17.76}{0.26} & \bms{12.82}{0.21} & \bms{17.76}{0.32} & \bms{12.58}{0.35} & \bms{17.51}{0.26} & \bms{12.73}{0.23} & \bms{17.62}{0.53} \\
		\hline
		\end{tabular}
            }
           \vspace{-10pt}
\end{table*}

\subsection{Efficiency Analysis}\label{sec-effi}

\begin{table}[h]
    \vspace{-25pt}
    \centering
    \betweenTinyAndScript
    \caption{Efficiency and complexity comparison.}
    \label{tab:efficiency}
    \begin{tabular}{c|c|c|c||c|c|c}
        \hline
        \multirow{2}{*}{Model} 
        & \multicolumn{3}{c||}{\textit{KuaiRand-Pure}} 
        & \multicolumn{3}{c}{\textit{Alimama}} \\ \cline{2-7}
        & Training time (s) & Params (MB) & GFLOPs 
        & Training time (s) & Params (MB) & GFLOPs \\ \hline
        ADIN     & \ms{8352}{1082} & 7.70 & 2.82  & \ms{5638}{355} & 39.56 & 4.22 \\ \hline
        SASS     & \ms{8745}{1028} & 3.27 & 10.19 & \ms{3977}{179} & 40.10 & 10.89 \\ \hline
        M5       & \ms{10080}{1879} & 13.59 & 43.87 & \ms{3861}{180} & 48.33 & 53.17 \\ \hline
        PERSCEN  & \ms{13431}{2174} & 4.30 & 8.52  & \ms{4166}{217} & 39.80 & 12.84 \\ \hline
        \textbf{\method{}} & \ms{7174}{976} & 2.11 & 6.21  & \ms{4135}{312} & 38.96 & 5.51 \\ \hline
    \end{tabular}
    \vspace{-15pt}
\end{table}

As shown in Table~\ref{tab:efficiency}, \method{} consistently outperforms strong multi-scenario baselines while maintaining high efficiency. On both KuaiRand-Pure and Alimama, it achieves competitive training speed with lower parameter and computational complexity, confirming that its lightweight design delivers strong performance without incurring significant overhead, making it well-suited for industrial recommender systems.

\subsection{Ablation Study}

To verify the effectiveness of each key component in the \method{} model, we conducted a series of ablation studies on the KuaiRand-Pure dataset, with the results presented in Table~\ref{tab:ablation1}.

\begin{table}[t]
\vspace{-5pt}
\caption{Ablation study on KuaiRand-Pure.}
\setlength\tabcolsep{1pt}
\label{tab:ablation1}
\resizebox{\textwidth}{!}{
    \centering
    \betweenTinyAndScript
 \vspace{-8px}
\begin{tabular}{c|cc|cc|cc|cc}
    \hline
    \multirow{2}{*}{Variant} & \multicolumn{2}{c|}{K1~(84\%)} & \multicolumn{2}{c|}{K2~(9\%)} & \multicolumn{2}{c|}{K3~(4\%)} & \multicolumn{2}{c}{K4~(3\%)} \\  
    & R@500(\%) & R@1000(\%) & R@500(\%) & R@1000(\%) & R@500(\%) & R@1000(\%) & R@500(\%) & R@1000(\%) \\   
    \hline
    w/o SAP 
    & \ms{18.08}{0.63} & \ms{29.05}{0.83} 
    & \ms{29.05}{0.30} & \ms{40.68}{0.47} 
    & \ms{19.63}{1.45} & \ms{29.93}{0.97} 
    & \ms{18.46}{1.24} & \ms{27.60}{1.65}\\
    \hline
    w/o Distillation 
    & \ms{18.74}{0.38} & \ms{29.65}{0.22} 
    & \ms{29.04}{0.53} & \ms{40.27}{0.76} 
    & \ms{19.27}{0.25} & \ms{29.28}{0.44} 
    & \ms{18.80}{0.69} & \ms{28.52}{0.97} \\
    \hline
    w/o DSBN 
    & \ms{18.53}{0.14} & \ms{29.49}{0.15} 
    & \ms{30.50}{0.90} & \ms{41.55}{0.82} 
    & \ms{18.66}{0.69} & \ms{29.88}{0.55} 
    & \ms{20.27}{0.91} & \ms{30.55}{0.78} \\
    \hline
    \method{} 
    & \bms{18.86}{0.24} & \bms{30.06}{0.23} 
    & \bms{31.04}{0.32} & \bms{42.66}{0.18} 
    & \bms{20.26}{0.75} & \bms{30.61}{0.78} 
    & \bms{21.87}{0.72} & \bms{32.60}{0.64} \\
    \hline
\end{tabular}
}
\vspace{-10pt}
\end{table}

\begin{itemize}
	\item \textbf{w/o SAP} Removing the core Scenario-Adaptive Projection (SAP) module led to a significant performance degradation across all scenarios, particularly in the data-sparse scenarios. This demonstrates the crucial role of the SAP module in dynamically adapting to different scenarios and mitigating the data imbalance issue.
        \item \textbf{w/o Distillation} Removing the knowledge distillation framework and training only with the task loss caused a clear performance drop. Distillation allows the student model to inherit richer interaction knowledge from a teacher with joint user-item featurest. The degradation without it highlights its effectiveness in overcoming the two-tower model’s structural limitations.
        \item \textbf{w/o DSBN} The removal of Domain-Specific Batch Normalization (DSBN) also led to a decline in performance. This indicates that employing independent normalization parameters for different scenarios helps the model learn more discriminative scenario-specific representations.
    
\end{itemize}

In summary, the ablation results clearly demonstrate that the SAP module, the knowledge distillation framework, and DSBN are all indispensable components of the \method{} model, working collaboratively to achieve its superior performance.

\subsection{Hyperparameter Sensitivity Analysis}
\label{app:hyper}
\subsubsection{Number of Experts K}
\begin{figure}[h]
	\vspace{-10pt}
	\centering
	\subfigure[Scenario K1]{
		\includegraphics[width=0.225\textwidth]{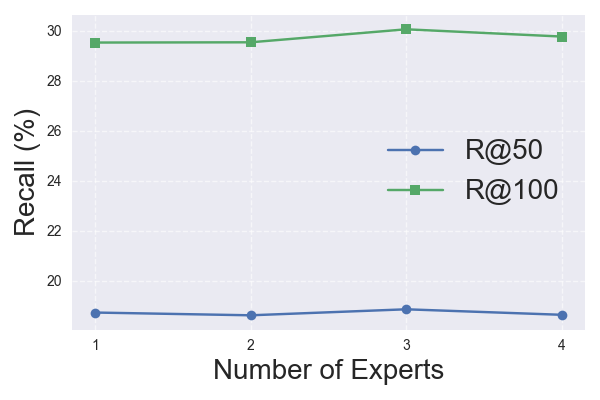}
	}
	\subfigure[Scenario K2]{
		\includegraphics[width=0.225\textwidth]{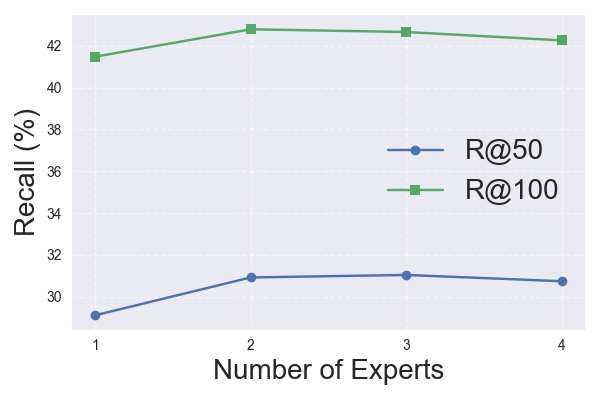}
	}
	\subfigure[Scenario K3]{
		\includegraphics[width=0.225\textwidth]{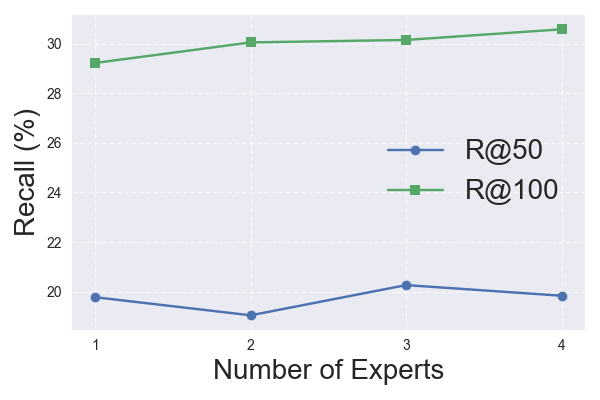}
	}
	\subfigure[Scenario K4]{
		\includegraphics[width=0.225\textwidth]{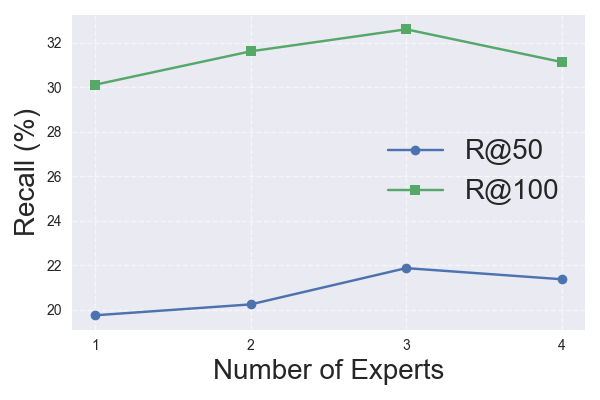}
	}
	\caption{Varying the number of experts for \method{} in KuaiRand-Pure.}
	\vspace{-30pt}
	\label{fig:expertanalysis}
\end{figure}

We examined the effect of expert number $K$ on \method{} performance (Figure~\ref{fig:expertanalysis}). Increasing $K$ generally improves results by capturing richer patterns, but gains plateau and may decline once $K$ becomes too large due to overfitting and redundancy. Balancing accuracy and efficiency, we chose an appropriate $K=3$ for our experiments.

\subsubsection{Rank R}

\begin{figure}[t]
	\centering
	\subfigure[Scenario K1]{
		\includegraphics[width=0.225\textwidth]{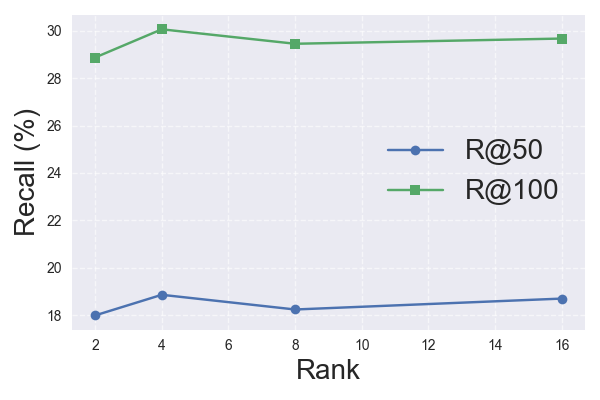}
	}
	\subfigure[Scenario K2]{
		\includegraphics[width=0.225\textwidth]{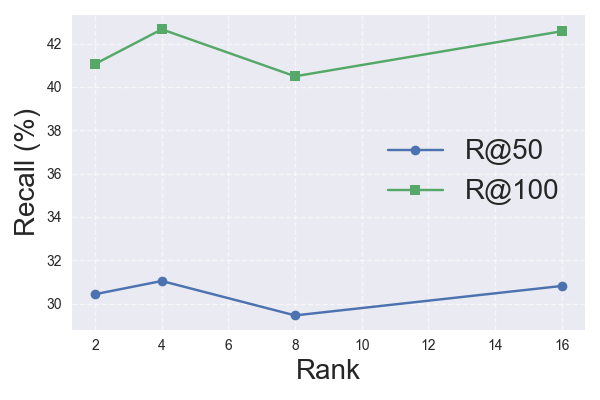}
	}
	\subfigure[Scenario K3]{
		\includegraphics[width=0.225\textwidth]{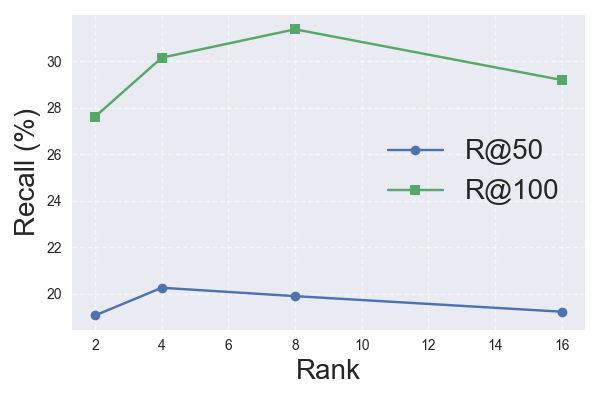}
	}
	\subfigure[Scenario K4]{
		\includegraphics[width=0.225\textwidth]{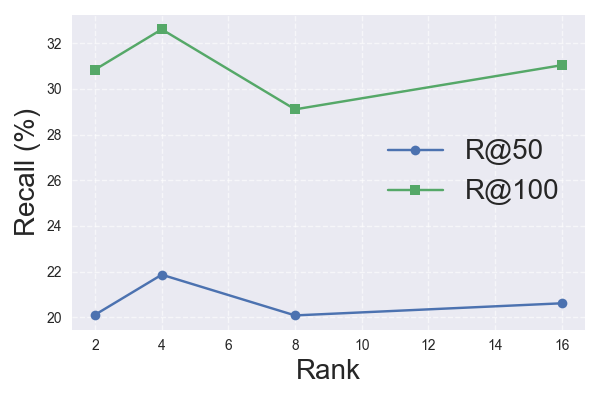}
	}
	\caption{Varying the number of rank R for \method{} in KuaiRand-Pure.}
	\vspace{-20pt}
	\label{fig:rankanalysis}
\end{figure}
We analyzed the sensitivity of SAP to the rank $R$ of its low-rank decomposition (Figure~\ref{fig:rankanalysis}). Increasing $R$ initially improves performance by capturing richer scenario-specific information, but gains soon saturate. A relatively small $R$ is thus sufficient to provide effective parameters while preserving the lightweight nature of the module.

\subsection{Case Study}\label{sec:case}

\begin{figure}[b]
	\centering
	\includegraphics[width=0.65\textwidth]{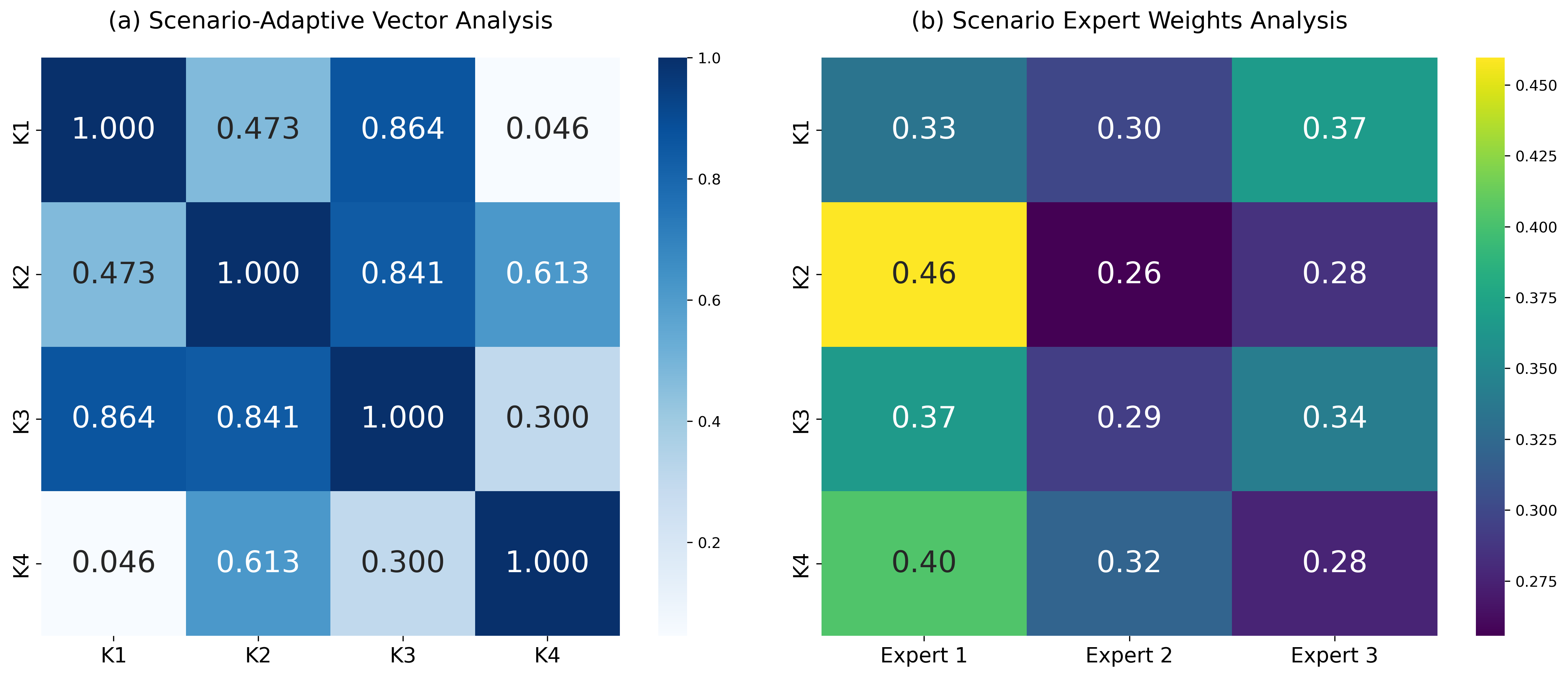}
	\caption{Case study on the KuaiRand-Pure dataset. (a) Cosine similarity of scenario-adaptive vectors $b_s$ from the SAP module. (b) Average expert activation weights from the gating network, showing clear specialization across scenarios.}
	\vspace{-10pt}
	\label{fig:case_study}
\end{figure}

To gain a deeper qualitative understanding of our proposed \method{} framework, we conduct a case study to visualize the behavior of its two core components: the Scenario-Adaptive Projection (SAP) module and the multi-expert layer. The analysis is performed on the KuaiRand-Pure dataset, and the results are presented in Figure~\ref{fig:case_study}.

\textbf{Analysis of Scenario-Adaptive Projections.} 
Figure~\ref{fig:case_study}(a) shows the cosine similarity of scenario-adaptive vectors from the SAP module. The distinct vector for the largest versus the most data-sparse scenario indicates that SAP generates specialized projections for tail scenarios, while the high similarity among certain scenarios suggests it can also capture shared characteristics when appropriate.

\textbf{Analysis of Expert Specialization.} 
Figure~\ref{fig:case_study}(b) visualizes the average activation weights of the three experts as determined by the gating network. The results reveal a clear division of labor: one expert focuses on the main high-traffic scenario, another specializes in smaller scenarios, while the third remains relatively balanced. This demonstrates that the MMOE architecture effectively allocates experts to capture diverse scenario patterns, which is crucial for balanced performance.



\section{Conclusion}
In this work, we introduced DSMOE to address the structural and distributional bottlenecks of applying MMOE to matching. By integrating Scenario-Adaptive Projection (SAP) and interaction-aware distillation, we effectively reconciled the conflict between the independent encoding required for efficiency and the feature interaction required for accuracy. Beyond superior metrics, DSMOE offers a generalizable paradigm for interaction-consistent retrieval: proving that the trade-off between inference speed and interaction complexity is not zero-sum, but can be harmonized via architectural decoupling and supervisory transfer. Our future roadmap involves exploring real-world multi-modal and multi-scene scenarios. By leveraging advances in cross-modal learning~\cite{li2025survey,qiao2025improving}, we seek to construct more robust recommendation systems for diverse application fields.

\bibliographystyle{splncs04} 
\bibliography{ref}




\end{document}